\documentclass[%
reprint,
 amsmath,amssymb,
 aps,prl
]{revtex4-1}

\usepackage[utf8]{inputenc}
\usepackage[english]{babel}
\usepackage{graphicx}
\usepackage{dcolumn}
\usepackage{bm}
\usepackage{braket}
\usepackage{color}

\newcommand{\im}{\operatorname{i}}

\DeclareMathOperator*{\Tr}{Tr}
\allowdisplaybreaks

\begin{document}

\preprint{APS/123-QED}

\title{Melting Si: beyond density functional theory}
\author{Florian Dorner}
\author{Zoran Sukurma}
\author{Christoph Dellago}
\author{Georg Kresse}
\email{georg.kresse@univie.ac.at}
\affiliation{%
 University of Vienna, Faculty of Physics and Center for Computational Materials
Sciences, Sensengasse 8/12, 1090 Wien
}%

\date{\today}

\begin{abstract}
The melting point of silicon  in the cubic diamond phase is calculated using the random phase approximation (RPA).
The RPA  includes exact exchange as well as an  approximate treatment of local as well
as non-local many body correlation effects of the electrons. We predict a melting temperature of about 1735~K and 1640~K
without and with core polarization effects, respectively. 
Both values are within 3~\% of the experimental melting temperature of 1687~K. 
In comparison, the commonly used
gradient approximation to density functional theory predicts a melting point that is 200~K too low, and hybrid functionals
overestimate the melting point by 150~K. We correlate the predicted melting point
with the energy difference between cubic diamond and the beta-tin phase of silicon, establishing 
that this energy difference is an important benchmark for the development of approximate functionals.
The current results establish that the RPA can be used to predict accurate finite temperature properties
and underlines the excellent predictive properties of the RPA for condensed matter.
\end{abstract}

\pacs{Valid PACS appear here}
\maketitle


An accurate prediction of the melting point of solids using first principles methods is still among the most challenging tasks 
in materials modeling. Such calculations are computationally demanding since in order to predict
accurate melting temperatures, the free energy of the solid and the liquid must be calculated with very high
precision, typically to a tolerance of about 1~meV per atom for 10~K precision~\cite{morawietz2016van}.
This requires stringent convergence
tests as well as carefully laid out procedures in order to obtain sufficiently accurate absolute energies. 
Alternatives to a separate evaluation of the free energy of the liquid and solid exist, but generally these require one
to consider very large supercells with on the order of 1000 atoms modeling directly the two-phase coexistence~\cite{morris1994melting,alfe2003Almeltingcoexist,Li_PRLett.104.185701,pedersen2013direct,Hummel2013interfacepinning,morawietz2016van}.
These methods are not only very expensive, but also finite size error estimates are difficult to estimate  using first principles techniques~\cite{alfe2003Almeltingcoexist,Al_PRB.80.094102,Li_PRLett.104.185701}.

There is a second, maybe even more important issue affecting the prediction of melting temperatures. Typically, 
the presently available density functionals
are not accurate enough to yield reliable predictions for the melting point. A prime example
is the melting of silicon, which the local density approximation predicts to occur between 1300-1350~K~\cite{SuginoCar1995Simelting,alfe2003lSimelting}, almost  20\% below the experimental melting point at 1687~K. 
Although the gradient approximation PBE (Perdew-Burke-Ernzerhof)~\cite{Perdew_PBE_1996} improves upon 
this value increasing the melting point to about 1480~K~\cite{alfe2003lSimelting}, most yet published predictions underestimate the melting point of Si by at least 10~\%. 
As we will show below, the hybrid functional HSE06 (Heyd-Scuseria-Ernzerhof)~\cite{HSE06} and the recently
proposed  SCAN (Strongly Constrained Appropriately Normed) functional~\cite{Sun_SCAN_2015} perform slightly better, but
overestimate the melting point by 120 and 150~K, respectively. 
One, therefore certainly needs to go beyond semi-local and hybrid functionals, as for instance
done for Fe based on diffusion Monte-Carlo calculations\cite{Fe_PRL.103.078501}. This and laying out our precise
procedures is the main goal of the present letter. 

Central to the calculation of free energies is that the free energy difference $ F_1-F_0$ between two Hamiltonians
with potential energies $U_0({\bf R})$ and $U_1({\bf R})$ depending on positions $\bf R$ 
can be calculated either by thermodynamic perturbation 
theory (TPT)
\begin{equation}
\label{equ:TP}
  F_1- F_0 = -\frac{1}{\beta} \ln \langle  e^{-\beta [U_1({\bf R}) - U_0({\bf R})]} \rangle_0  
\end{equation}
or by a thermodynamic integration (TI)
\begin{equation}
\label{equ:coupling}
  F_1-F_0 = \int_0^1 d \lambda \langle U_1({\bf R}) -U_0({\bf R}) \rangle_\lambda.
\end{equation}
Here the notation $\langle A({\bf R}) \rangle_\lambda$ implies that the expectation value of $A$ is
evaluated for the ensemble corresponding to the classical Hamiltonian 
\[
H_\lambda({\bf R},{\bf P})= (1-\lambda)  \,U_0({\bf R}) + \lambda \, U_1({\bf R}) +{\bf P}^2/ (2M), 
\]
where $\bf P$ are the momenta of the atoms and $M$ is their atomic mass.
Non-adiabatic switching and combinations of both methods are possible as well~\cite{jarzynski1997nonequilibrium}, but are not necessarily computationally more efficient
than the two standard procedures \cite{oberhofer2005biased}.
If first principles calculations are used, one usually determines $U(\bf R)$ by the adiabatic Born-Oppenheimer approximation and density functional theory (DFT), {\em i.e.}, for each set of atomic
positions $\bf R$, the energy of the electronic degrees of freedom is minimized and the potential energy $U({\bf R})$, 
which then  depends parametrically
on the atomic positions $\bf R$, is calculated. 

Here, we will use a computationally much more involved many body description for the electrons, {\em i.e.}  the potential energy as a function of the atomic positions 
$U(\bf R)$ is calculated using the random phase approximation (RPA) to the electronic correlation energy. 
In this approximation,  first a DFT calculation is performed using an approximate density functional (here the PBE functional)~\cite{Perdew_PBE_1996}.
Then the exact exchange and RPA correlation energy are evaluated as
\cite{Nozier_RPA_1958,Langreth_RPA_1977,Miyake_RPA_2002,Fuchs_RPA_2005,Furche2008,Harl_PRL_RPA_2009,Schimka_Nat_RPA_2010}:
 \begin{equation}
  U_{\rm RPA} = E_{\rm EXX} + \frac{1}{2 \pi} \int_0^\infty {\rm d} \nu \Tr[ \ln(1-\chi(\im\nu)\operatorname{v})+\chi(\im \nu)\operatorname{v}], \label{equ:RPA_energy}
 \end{equation}
where $E_{\rm EXX}$ is the Hartree-Fock energy functional evaluated for PBE orbitals, $\chi(\im\nu)$  is the independent particle  polarizability calculated using PBE  orbitals and one electron energies, and $\operatorname{v}$
is the Coulomb kernel.
It has been shown that this approximation  describes  diverse bonds 
from covalent, over ionic, to  van der Waals like more accurately than available DFT functionals~\cite{Harl_PRL_RPA_2009,Lebegue2010,Schimka_Nat_RPA_2010,Torres_PRM.1.060803}.
Although we have recently presented a method to calculate interatomic forces in the
RPA~\cite{ramberger2016analytic}, finite temperature molecular dynamics (MD) simulations
for Si are still too demanding to allow one to use TI. 

\begin{figure}
    \begin{center}
   \includegraphics[width=55mm,clip=true]{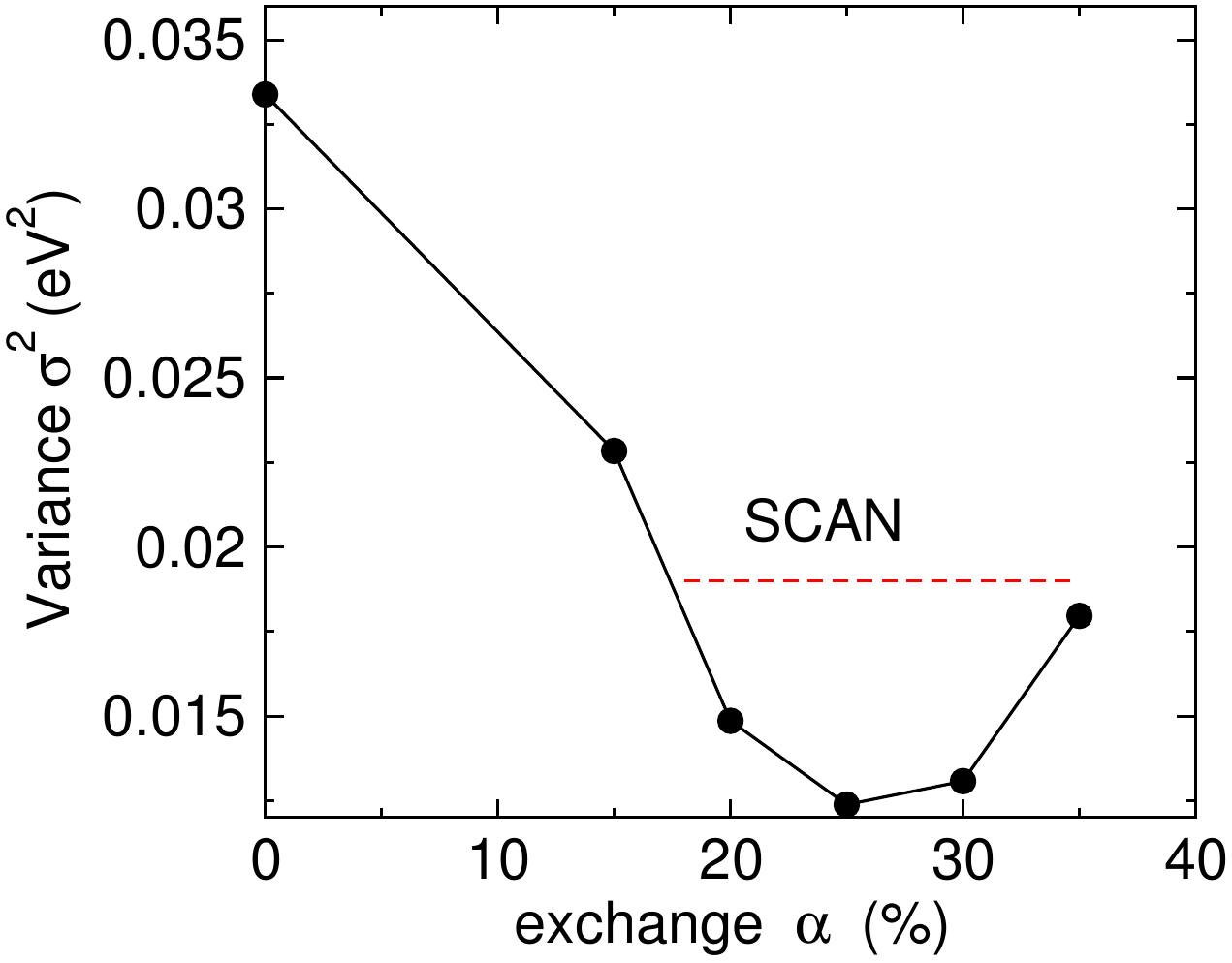}
    \end{center}
   \caption{
   Variance of the energy difference between the RPA and 
   hybrid functional calculations $\sigma^2 = {\rm Var}(U_{\rm RPA}-U_{\rm DFT})$ as a function of the amount $\alpha$ of exact exchange.
 The ensembles were created using the corresponding hybrid functional. The variance was calculated for 200 configurations
 picked from the MD  trajectory.
}
\label{fig:variance}
\end{figure}

Hence, we have to rely on TPT and we now first explain how we determine the free energy of the {\em liquid}. 
The obvious procedure to do this  is to use DFT generated ensembles ($U_0=U_{\rm DFT}$), calculate their  free energy 
and perform TPT to the RPA ($U_1=U_{\rm RPA}$). 
To minimize the statistical error bars  in TPT, one needs to minimize the variance of the  
exponential function in Eq. (\ref{equ:TP}).
This requires one to make a judicious choice for the reference DFT potential energy $U_{\rm DFT}$. 
To guide our choice, we have hence performed
MD simulations for liquid Si at 1687~K using 32 atoms, 2 k-points and various functionals, 
including hybrid functionals, where the amount of the exact
exchange $\alpha$ was varied between $\alpha=0.0$ (recovering the PBE functional) and $\alpha=0.35$ (at $\alpha=0.25$ the HSE06 functional is recovered) ~\cite{HSE06}. 
We then calculated the energy difference $U_{\rm RPA}-U_{\rm DFT}$
for 200 uncorrelated configurations picked from the DFT-MD trajectory and determined the variance of
the energy differences (see Fig. \ref{fig:variance})\cite{bokdam2017assessing}. This
assumes that the variance is sufficiently small to approximate the exponential in Eq. (\ref{equ:TP})
by its first order Taylor expansion around the mean energy shift.
Remarkably, with the standard choice of HF exchange $\alpha=0.25$, the variance between the hybrid functional 
and the RPA is smallest. In other words, the HSE06 functional generates an ensemble that is most similar
to the RPA and constitutes the best choice to perform TPT to the RPA. Since hybrid functional calculations are still rather expensive, we also investigated semi-local and meta-GGA 
functionals to determine the best ``cheap'' semi-local functional. We found that the SCAN functional also yields an excellent description of
the ``RPA-liquid'' with a variance only about twice as large as for HSE06 (see Fig. \ref{fig:variance}). 

This settles that we should use HSE06 and SCAN for our DFT calculations,  and we added PBE to compare to previous calculations.
To evaluate the free energy difference between the liquid and solid for these functionals, we follow similar strategies as 
previously employed by de Wijs, Kresse and Gillan~\cite{deWijs1998Almelting},
later refined by Alfe and Gillan~\cite{alfe2003lSimelting}. 

In the first step of our calculations, the {\em equilibrium volume} of the solid and liquid was calculated
at finite temperature for each considered DFT functional (compare Tab. \ref{tab:summary}).
To this end, $NVT$ MD simulations  
for at least three volumes $V$ were performed and the instantaneous stress tensor $\sigma=\partial U({\bf R}) / \partial \epsilon $ \cite{nielsen1983first}
at each time step (as determined by VASP) was recorded and  averaged to obtain the 
macroscopic stress tensor $\bar {\bf \sigma}$
\[
\bar {\bf \sigma} = - \frac{\partial F}{\partial {\bf \epsilon}} = \frac{1}{\beta}\frac{\partial \log Z }{\partial {\bf \epsilon}}= \Big \langle \frac{\partial \,U({\bf R}) }{\partial {\bf \epsilon}} \Big \rangle + \mbox{ideal gas term},
\]
where ${\bf \epsilon}$ is the strain tensors.
The finite temperature equilibrium volume was determined as the volume where the trace of $ \bar  {\bf\sigma}$ 
is zero. 

To obtain the {\em free energy of the solid}, we first calculated the harmonic vibrational
frequencies using a finite supercell. 
The phonon dispersion relation and the harmonic free energy are determined using a very fine grid of Bloch
wave vectors  by assuming that the force constants are  zero beyond the interaction range of the supercell \cite{kresse1995ab}.
To be compatible with the classical MD simulations performed for the liquid, we
used Maxwell-Boltzmann statistics instead of Bose statistics to determine the free energy of
the harmonic oscillations, although the difference between the two statistics is smaller than 1~meV
at $T=$1687~K. 64 atom supercells and  $2\times 2\times 2$ k-points are found to be sufficient to obtain a free energy  converged to 
better than 0.5~meV, if  the vibrational frequencies are Fourier-interpolated
to a very dense grid of  wave vectors as explained above (see supplementary). 

The anharmonic contributions are calculated by TI
from the harmonic case to the full {\em ab initio} Hamiltonian. To obtain accurate integrals, we used
a three-point Simpson or three-point Gauss quadrature for $\lambda$ in Eq. (\ref{equ:coupling}), 64 atoms and $2\times 2\times 2$ k-points. TPT
was then used to integrate from $2\times 2 \times 2$ k-points to $3\times 3 \times 3$ k-points
(such an up-sampling of k-points using TPT has previously been referred to as UP-TILD~\cite{Grabowski2009,FreysoldtRMP2014}).
Throughout this work, we use the second order cumulant expansion of Eq. (\ref{equ:TP}) for the TPT~\cite{zwanzig1954rw}:
\begin{equation}
\label{equ:TP2}
  F_1- F_0 \approx  \langle \Delta U  \rangle_0 - \frac{\beta}{2} \langle (\Delta U  - \langle \Delta U \rangle)^2  \rangle_0, \quad \Delta U = U_1 - U_0.
\end{equation}
This approximation is {\em exact} if all cumulants of order higher than two vanish, which is the case if and only if the probability density $P_0(\Delta U)$ in the $\lambda=0$ ensemble is Gaussian. As we discuss in the supplementary, this condition is equivalent to the requirement that the integrand in Eq. (\ref{equ:coupling}) is a linear function of the coupling parameter $\lambda$. The approximation of Equ. (\ref{equ:TP2}) has the advantage of a much smaller statistical error than the full TPT equation.

\begin{figure}
    \begin{center}
   \includegraphics[width=42mm,clip=true]{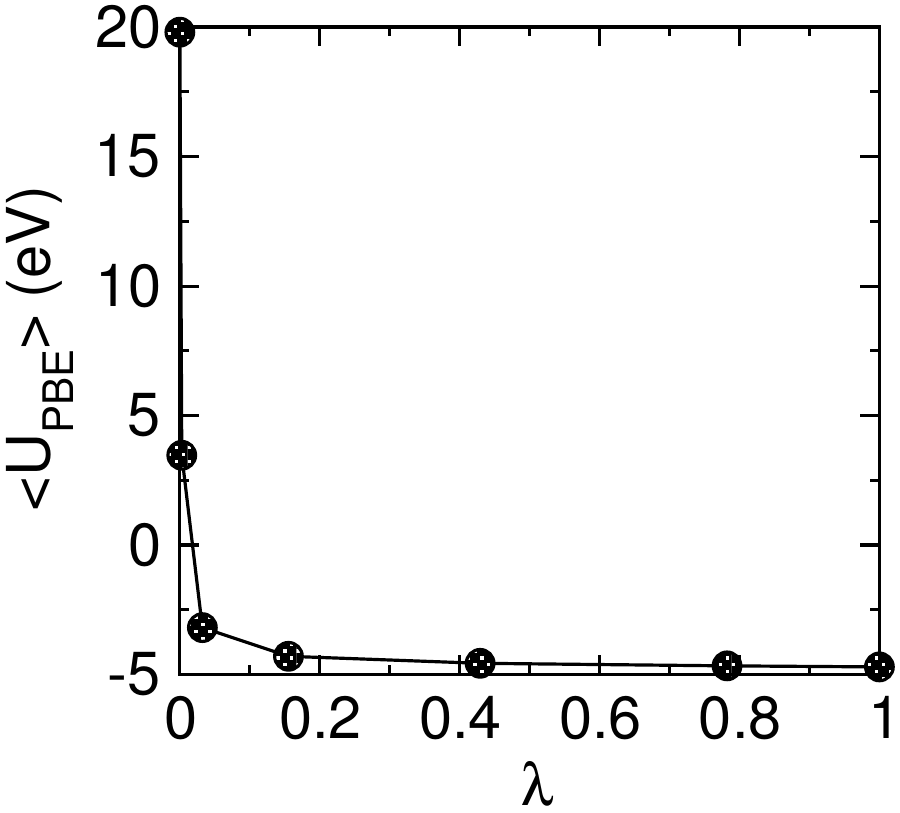}
   \includegraphics[width=42mm,clip=true]{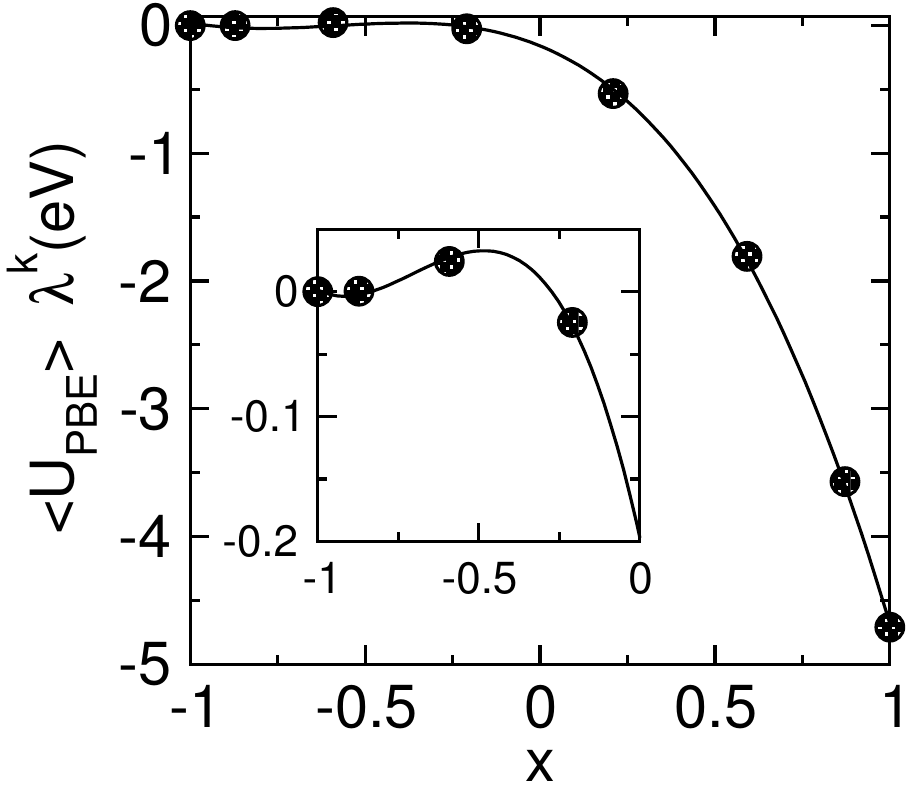}
    \end{center}
   \caption{
  Integrand of the TI from the ideal gas to liquid Si using SCAN 
  as a function of $\lambda$ and, after the transformation, as a function $x$ for $k=0.8$.
  The inset in the right panel shows a zoom in for negative $x$. The smooth curve indicates 
  the absence of any  phase-transitions.
}
\label{fig:histo}
\end{figure}

Calculations for the liquid are  less straightforward. The common practice is to integrate from a simple
classical force field $U_0$ to the first principles energy functional $U_1$, which
is inconvenient as it requires to  interface the electronic structure code to a force field code. Furthermore, for a covalent liquid
such as liquid Si with six nearest neighbors, it is also not a simple matter to find a suitably simple but accurate
force field. Instead, we decided to perform a direct TI
from the ideal gas to the full DFT Hamiltonian at fixed volume at the temperatures and volumes
summarized in Tab. \ref{tab:summary}. To allow for efficient calculations, the
first principles calculations are performed using the standard PBE functional and employ the $\Gamma$ point only,
and another TI  is then performed from the PBE functional to many k-points and
the desired DFT functional (PBE, SCAN or  HSE06). There are several intricacies that need to be addressed
for the first TI (ideal gas $\to$ PBE). The required coupling integral is given by
\begin{equation}
 \label{equ:flambda}
 \int_0^1 d \lambda \langle  U_{\rm PBE} \rangle_\lambda,
\end{equation}
where the classical Hamiltonian used to generate the ensembles is given by $H_\lambda = \lambda \, U_{\rm PBE}({\bf R }) + {\bf P}^2/(2M)$.
At small $\lambda$, the integrand becomes very large, since the atoms move as ideal gas particles, and whenever they approach 
each other the energy becomes hugely positive due to Pauli repulsion. On the other hand, for $\lambda>0.1$ the integrand is smooth
and well behaved. To deal with this issue, we perform a variable transformation from $\lambda$ to $x$: 
$\lambda(x) = (\frac{x+1}{2})^{\frac{1}{1-k}}$,
which maps the integration variable from $\lambda \in [0,1]$ to $x \in[-1,1]$. The integral (\ref{equ:flambda}) then becomes:
\begin{equation}
 \label{equ:fx}
 \int_{-1}^{1} f(\lambda(x)) \frac{d \lambda}{d x} dx = \frac{1}{2 (1-k)}\int_{-1}^{1} f(\lambda(x)) \lambda(x)^k dx.
\end{equation}
The key point is to choose $k$ sufficiently close to 1 that the transformed integrand (r.h.s. of Eq. \ref{equ:fx})
becomes zero at $x=-1$. For Si, we found that this is observed if $k \ge 0.7$. Since the integrand becomes zero
at $x=-1$  and since inclusion of the point at full coupling is  convenient, we  used a 
Gauss-Lobatto integration (a Gauss like integration rule that includes the end-points $x=\pm1$).
Here we found that an 8 point Gauss-Lobatto rule and $k=0.8$ yield a precision better then 0.5~meV per atom. 
To perform stable and accurate calculations, the time step in the MD simulation 
must be reduced to 0.5~fs at small $\lambda$ values, where
the interaction potential becomes hard sphere like. If this is not done, the VASP code
eventually can not find the electronic groundstate, whenever the atoms come too close to each other.

In the second step, we performed a TI from the PBE $\Gamma$-point calculation
to the desired functional and $2\times 2 \times 2$ k-points. Three-point Simpson or three-point Gauss integrations
were typically used, although a simple mid point rule or trapezoidal rule also gave errors of about 1~meV (see supplementary). 
As for the solid, thermodynamic perturbation theory was used to determine the free energy of the desired
functional using $3\times 3 \times 3$ k-points. Calculations for the liquid were performed using 64 atom ensembles,
as well as 128 and 216 atom ensembles for SCAN (see supplementary).

\begin{table}[t]
\caption{
Contributions to the free energy $F$ of the solid and liquid
at $T$ calculated for 64 atom supercells.
Energies are in units of eV per atom. If applicable, the energy of 
cubic diamond Si at T=0~K and a volume of 20.2~\AA$^3$ was subtracted.
Statistical errors are reported in the supplementary. 
The melting temperature $T_m$ and melting enthalpy $\Delta H_{sl}$ are also reported (experimental estimates
for  $\Delta H_{sl}$ are 0.52 and  0.47 eV, see Ref. \onlinecite{SuginoCar1995Simelting}).
}
\label{tab:summary}
\begin{ruledtabular}
\begin{tabular}{lrrr}
DFT functional           &  HSE      &   SCAN      &    PBE  \\
T(K)                     & 1687~K    &  1687~K     &    1450~K     \\
\hline
solid-Si          \\
volume $V$               &20.49 \AA$^3$   &   20.38 \AA$^3$    &  20.83 \AA$^3$   \\
ideal crystal $3\times 3 \times 3$
                         &   0.0024  &   -0.0012   &   -.0018 \\ 
$\Delta$ harmonic        &  -0.5955  &   -0.6117   &   -.4764 \\ 
$\Delta$ anharmonic      &  -0.0150  &   -0.0005   &   -.0135 \\ 
$TS$                     &   1.067   &    1.070    &   0.878  \\ 
\hline                                                                                 
liquid-Si          \\                                                                  
volume                   &  18.30 \AA$^3$   &   18.53 \AA$^3$    &  18.33 \AA$^3$  \\
ideal-gas                &  -1.6638  &   -1.6634   & -1.3997  \\ 
PBE $\Gamma$             &  -0.6439  &   -0.6485   & -0.4225  \\ 
DFT $3\times 3 \times 3$ &  -0.5724  &   -0.5699   & -0.4919  \\ 
$TS        $             &   1.542   &    1.542    &   1.291   \\ 
\hline
$T_m$(K)                 & 1813$\pm$12 &  1842$\pm$10  &  1449$\pm$10 \\
$\Delta H_{sl}$          &   0.522   &    0.526    &   0.412 \\
\end{tabular}
\end{ruledtabular}
\end{table}


Table \ref{tab:summary} summarizes our results. 
A few comments are in place here. (i) For the solid, we calculated
the difference between $2\times 2 \times 2$ and $3\times 3 \times 3$  k-points
for at least 20 finite temperature configurations, and the shift is constant and 
identical to the energy shift of the ideal crystal, when the k-point mesh is increased. 
Hence we have accounted for this contribution in the ideal crystal term.
(ii) HSE results in larger vibrational frequencies reducing the harmonic free energy. Remarkably this 
is almost exactly canceled by the anharmonic term. (iii)
The results were checked by performing TPT from HSE to SCAN and vice versa
finding agreement to within 1~meV for the solid and 2~meV for the liquid,
which is within the estimated error bars (see supplementary).

The predicted melting temperatures are 1449$\pm$10~K, 1813$\pm$12~K and  1842$\pm10$~K, for PBE, HSE06,  
and SCAN, respectively. Here, the melting point was estimated using the relation (see supplementary)~\cite{morawietz2016van}
\begin{equation}
 \label{equ:estimate}
  T_{\rm melt} = T + ( F_{l} - F_{s}) / ( S_{l}- S_{s}),
\end{equation}
where $F_{l/s}$ and $S_{l/s}$ are the free energy and entropy of the solid and liquid, respectively, evaluated at $T$.
To test the accuracy of this relation, the SCAN calculations were  repeated at 1800~K, predicting a melting
point of 1834$\pm10$~K. This suggests  the linear estimate via 
Eq. (\ref{equ:estimate}) to be accurate within the estimated error bars.
As a further test, we repeated the liquid and the solid 
state calculations using SCAN for 128 and 216 atoms at 1800~K finding agreement
to within 0.2~meV and 1~meV for the solid and liquid, respectively (see supplementary),
thus our estimates are technically converged. A previous study using SCAN and the two-phase coexistence method estimated
a melting point of $1652\pm46$ K but found  a difference of 100 K between
224 and 432 atom ensembles relegating a detailed finite size analysis to future studies \cite{scSi_PhysRevB.97.140103}.
In comparison, our free energy calculations are less
affected by finite size errors.
We  conclude that  none of the DFT models are satisfactory, with PBE being particularly
unsatisfactory underestimating the melting point by almost 15~\%, followed by SCAN with an 
overestimation of about 10~\%.

The RPA improves on these results substantially and predicts melting points of $1735\pm15$~K
if core polarization effects via the Si $2s$/$2p$ electrons are not included, and $1640\pm15$~K if  they are included.
Both values are within 3~\% of the experimental value.
The melting enthalpy  is predicted to be $\Delta H_{sl}=0.49(1)~$eV inbetween the experimental values
of 0.47 and 0.52 \cite{SuginoCar1995Simelting}.
The RPA free energy was determined by performing a TPT from SCAN or HSE06 to the RPA,
using $2\times 2 \times 2$ k-points and 64 atoms. For crystalline (liquid) Si,
20 (60) configurations suffice to obtain 
errors below 0.5~meV. Furthermore, the difference between
the SCAN and HSE06 reference point is about 5~K for the melting temperature (above we report the average). 
We conclude that the RPA to the correlation energy
of the electrons yields an excellent description of the melting properties of 
silicon.

\begin{figure}
    \begin{center}
   \includegraphics[width=65mm,clip=true]{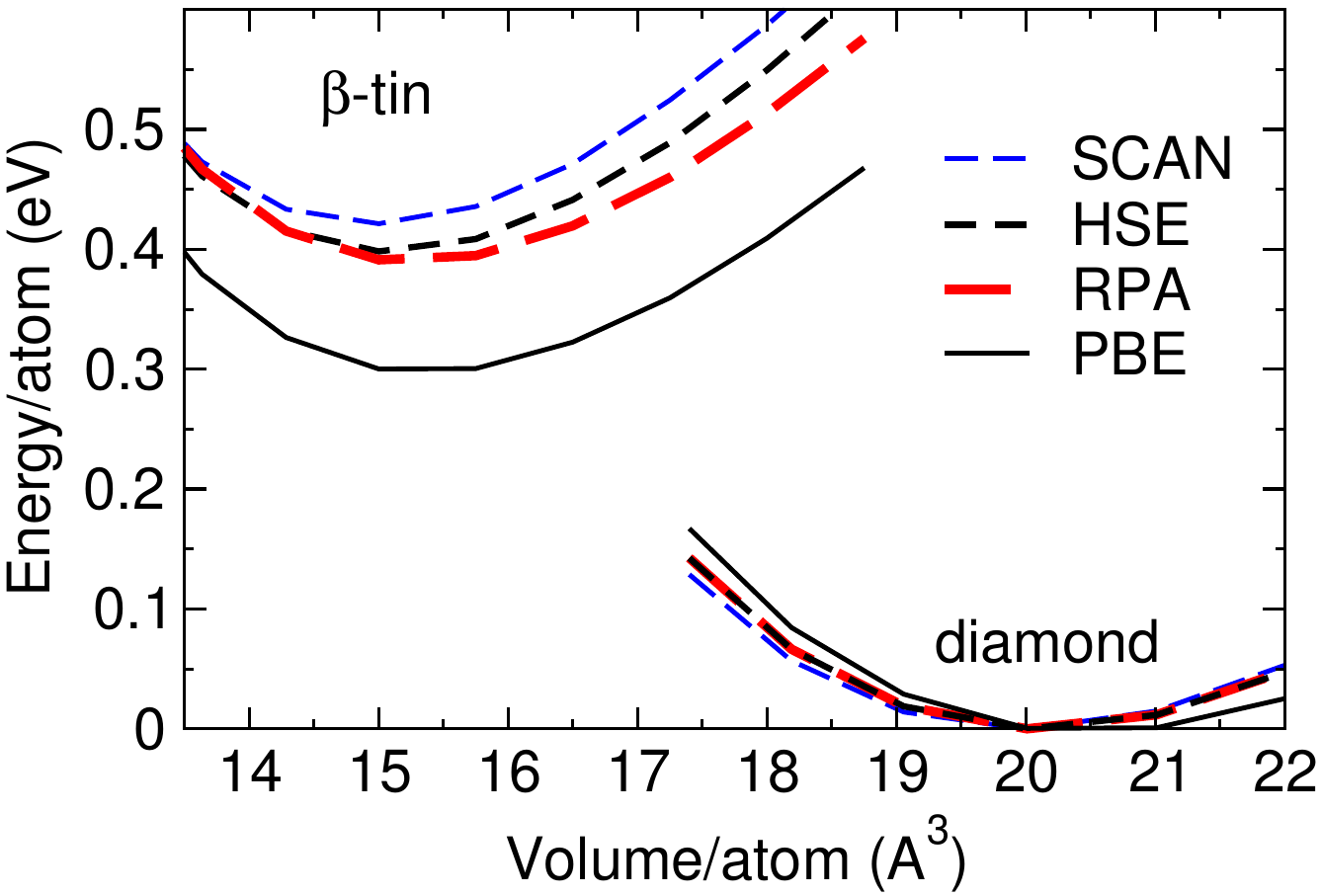}
    \end{center}
   \caption{
 Energy versus volume curves for cubic diamond Si and Si in the $\beta$-tin structure
for various functionals. For each functional the curves were
shifted vertically to align at zero for the cubic diamond phase. 
}
\label{fig:Siphase}
\end{figure}

Having established that the RPA is very accurate, we are
now exploring whether the melting point is related to the predicted energy
difference between the cubic diamond and $\beta$-tin phase of silicon as suggested in Ref. ~\onlinecite{alfe2003lSimelting}.
This argument rests on the observation that each Si atom has roughly 6 nearest neighbors
in the liquid, resembling the coordination in $\beta$-tin. 
In Fig. \ref{fig:Siphase} we show the energy-volume curves 
for the functionals considered here. Clearly, with respect to PBE, RPA, HSE and SCAN stabilize the 
tetrahedrally coordinated strucural motives compared to the six fold metallic ones \cite{Sun_SCAN_2016},
going in hand with increased melting temperatures.
The relation is, however, only qualitativ:
for instance, at $V=$18~\AA$^3$, SCAN shifts the $\beta$-tin structure
upwards by about 190 meV compared to PBE, which  considering Eq. (\ref{equ:estimate})
corresponds to a melting temperature change of about 650~K, twice
the predicted value. 

In summary, the RPA predicts very accurate
melting temperatures for silicon within few percent of the experiment.
This establishes again that, for condensed matter systems, the random phase
approximation  outperforms the available density functionals 
(including the recently suggested Strongly Constrained Appropriately Normed
functional).
By inspecting the variance of the energy difference between the random
phase approximation and various density functionals for many liquid
state configurations, we have also devised a strategy to pick the ``best'' available functional 
for the problem at hand.
This as well as the other procedures laid out here are straightforwardly applicable 
to other materials and hopefully pave the way towards accurate quantitative
melting point, and more generally, finite temperature calculations
for condensed matter using methods beyond density functional theory.

{\em Acknowledgment:} Funding by the Austrian Science Fund (FWF): F41 (SFB ViCoM) is grateful
acknowledged.  Computations were predominantly performed on the Vienna Scientific Cluster (VSC3).

\bibliography{literature}
\end{document}